\documentclass[a4paper,11pt]{article}
\pdfoutput=1 

\usepackage{jheppub} 

\usepackage[T1]{fontenc} 

\newcommand{\be}{\begin{equation}}
\newcommand{\ee}{\end{equation}}
\newcommand{\eqn}[1]{(\ref{#1})}

\title{\boldmath A Simple Holographic Insulator}

\author[]{Eric Mefford and}
\author[]{Gary T. Horowitz}

\affiliation[]{Department of Physics\\ University of California\\ Santa Barbara, CA 93106-4030}

\emailAdd{mefford@physics.ucsb.edu}
\emailAdd{gary@physics.ucsb.edu}

\abstract{We present a simple holographic model of an insulator.  Unlike most previous holographic insulators, the zero temperature infrared geometry is completely nonsingular. Both the low temperature DC conductivity and the optical conductivity at zero temperature satisfy power laws with the same exponent, given by the scaling dimension of an operator in the IR. Changing a parameter in the model converts it from an insulator to a conductor with a standard Drude peak.}

\begin{document} 
\maketitle
\flushbottom

\section{Introduction}
\label{sec:intro}

Gauge/gravity duality provides a new tool to study strongly correlated systems \cite{Hartnoll:2009sz,McGreevy:2009xe, Sachdev:2011wg}. In particular, it provides a novel way to study  states of matter at zero temperature. 
Indeed holographic models of superconductors \cite{Hartnoll:2008vx}, as well as conductors and insulators \cite{Hartnoll:2012rj,Donos:2012js,Donos:2013eha,Donos:2014uba,Gouteraux:2014hca,Ling:2014saa} have all been found, and some have properties similar to what is seen in exotic materials \cite{Donos:2012ra,Horowitz:2012ky}.

Previous discussions of holographic insulators fall into two classes. One is based on an asymptotically anti-de Sitter (AdS) solution called the AdS soliton. This solution has a gap for all excitations in the bulk and hence is dual to a gapped system \cite{Nishioka:2009zj,Horowitz:2010jq}. The other class starts with a system with finite charge density. In this case, translation invariance leads to momentum conservation which implies an infinite DC conductivity, $\sigma_{DC}$. (Charge carriers have no way to dissipate their momentum.) If one breaks translation invariance, either explicitly or spontaneously, the DC conductivity is finite. To obtain an insulator, one usually adds a perturbation which becomes large in the IR, leading to a bulk geometry which is singular in the interior. Since $T=0$, this is not a black hole singularity, but rather a timelike or null naked singularity.

In this note we show how to obtain a holographic description of an insulator using a nonsingular bulk geometry. Like the AdS soliton, we work at zero net charge density, so we can keep translation invariance and have finite $\sigma_{DC}$.\footnote{With an equal number of positive and negative charge carriers,  the charge current and momentum decouple since an applied electric field induces a current, but the net momentum stays zero.} However, unlike the approach based on the AdS soliton, at low temperature, the entropy scales like a power of $T$  showing the system is not gapped. The IR geometry is simply another AdS spacetime, so our solution describes a renormalization group flow from one CFT to another. We induce this flow by adding a relevant deformation to the original CFT. We will see that by tuning the interaction between a scalar field and gauge field, we can ensure that $\sigma_{DC} = 0$.

In a little more detail,  we construct our holographic insulator by starting with gravity coupled to a  scalar field $\psi$ with a  ``Mexican hat" type potential $V(\psi)$. By modifying the boundary conditions on the scalar in a way corresponding to a relevant double trace deformation, we induce the scalar to turn on at low temperature. The zero temperature solution is then a standard domain wall interpolating between the AdS corresponding to $\psi = 0 $ at infinity and the AdS corresponding to the minimum of the potential at  $\psi = \psi_c$ in the interior (see, e.g., \cite{Freedman:1999gp}). Finally, we add a Maxwell term  to the action with a coefficient $G(\psi)$. This function is chosen to vanish when $\psi = \psi_c$. Since it has been shown that the DC conductivity is simply given by the value of $G(\psi)$ on the horizon \cite{Iqbal:2008by}, it follows immediately that $\sigma_{DC}$ vanishes at zero temperature and we have an insulator.

We will show that both the DC conductivity at low temperature and the optical conductivity at zero temperature satisfy power laws:
\begin{equation}
{\rm Re}\ \sigma_{DC} \sim T^{2\Delta_\psi}\quad\quad\quad \lim_{T\to 0} {\rm Re}\ \sigma \sim \omega^{2\Delta_\psi}
\label{eq:intro}
\end{equation}
where the exponent $\Delta_\psi$ is simply related to the dimension of the operator dual to our scalar in the IR CFT. These results are similar to the behavior found in more complicated constructions of holographic insulators starting with a nonzero charge density \cite{Donos:2012js,Donos:2013eha}. However in those cases the power law is somewhat surprising given the singular nature of the IR geometry, and is the result of an approximate scaling symmetry in an intermediate regime. In contrast, the power law in our case is simply the result of the fact that our low energy theory has no scale.

The organization of this paper is as follows. We will start by introducing our model and discussing  how imposing a modified boundary condition for our scalar field can induce an instability which turns on the scalar field. (This corresponds to adding a relevant double-trace deformation of the CFT.) We will then discuss how to compute the conductivity and present both numerical and analytic arguments for the power laws. Finally we show that this same model with a slightly different $G(\psi)$ can also describe a conductor with a standard Drude peak.


\section{The Model}
\label{sec:ourmodel}

\begin{figure}[tbp] 
  \centering
\includegraphics[width=0.60\textwidth]{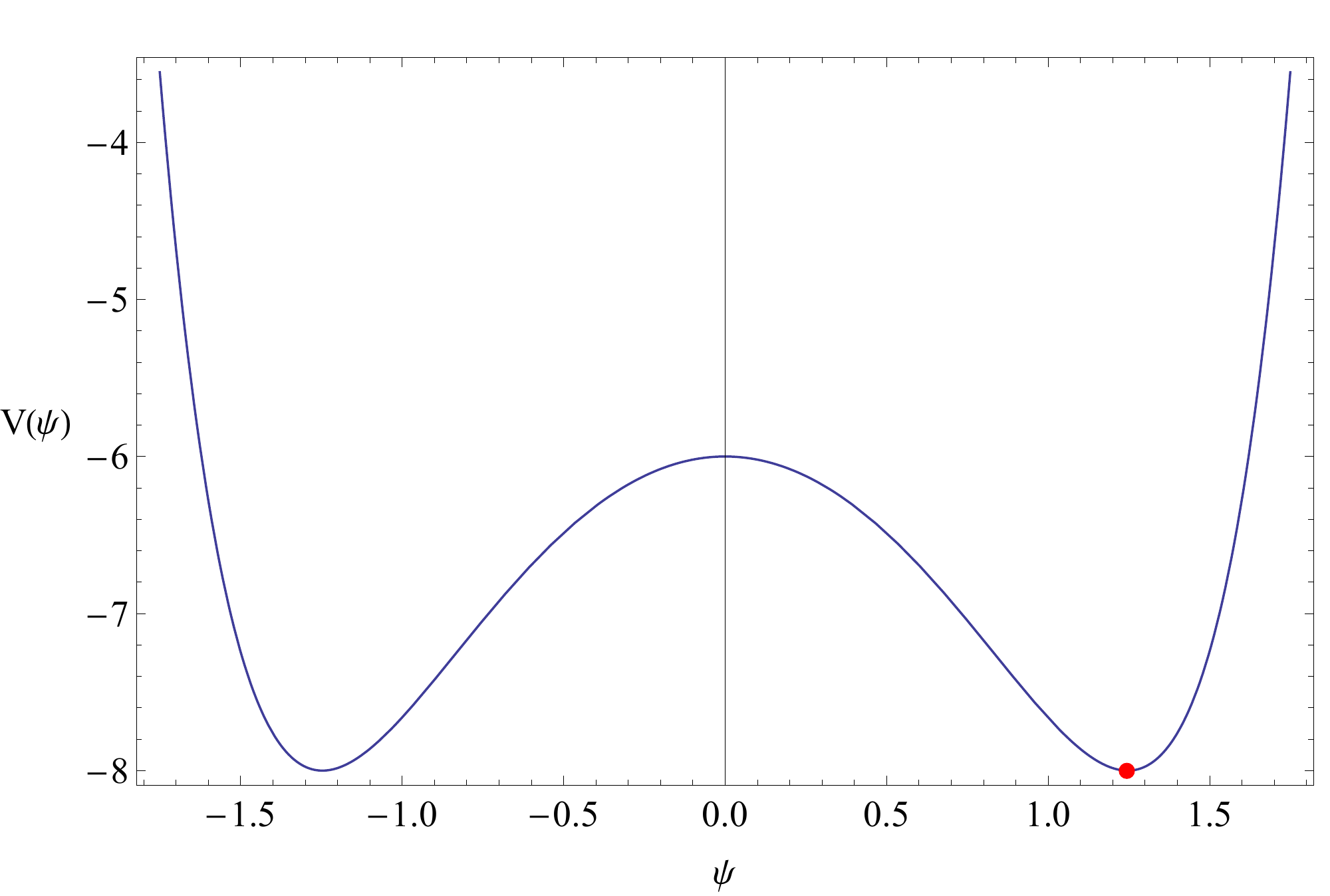}
  \caption{\label{fig:p} A plot of our potential. The minimum is at $\psi_c = \sqrt{2}\ln(1+\sqrt{2})$ with minimum value $V(\psi_c) = - 8$.}
\end{figure}

We will study a $3+1$ dimensional gravitational theory in anti-de Sitter spacetime with a real scalar field $\psi$ and a $U(1)$ gauge field, $A_\mu$. These are dual to a $2+1$ dimensional CFT with a scalar operator $\mathcal{O}$ and a conserved current $J^{\mu}$, respectively.  (The model is easily extended to other dimensions.) The action for these fields is
\begin{equation}
S = \int d^4 x \sqrt{-g} \left[R-\frac{1}{4}G(\psi)F^2 - (\nabla\psi)^2 - V(\psi)\right]
\label{eq:action}
\end{equation}
where
\begin{equation}
G(\psi) = (1+g\psi^2)^2, \quad\quad\quad V(\psi) = -{6\over L^2} +\frac{1}{L^2} \sinh^2(\psi/\sqrt{2})\left [\cosh(\sqrt{2}\psi)-5\right ].
\label{eq:gpsi}
\end{equation}
 The particular form of $V(\psi)$ is not important.  All we need is that it has a local maximum at $\psi=0$ (with $m^2$ within a suitable range discussed below),  and a global minimum at some nonzero value $\psi_c$.\footnote{For stability of the gravity solution, we also require that  $V$ can be derived from a certain superpotential, as we will discuss shortly.}
  The particular choice we have made  comes from a consistent supergravity truncation~\cite{Gubser:2009gp} and is shown in Fig. \ref{fig:p}. The particular form of $G(\psi)$ is also not crucial. What we need to model an insulator is a positive function that vanishes at $\psi_c$.  This will hold with the form of $G$ that we have chosen if we set $g = -1/\psi_c^2$.
We will see later that this same theory will describe a conductor with standard Drude peak, if we  take $ g>0$.

We make an ansatz for an asymptotically Poincar\'e $AdS_4$ metric,
\begin{equation}
ds^2 = -f(r)dt^2 + \frac{dr^2}{f(r)} + h(r)^2 d\vec{x}^{\;2}
\end{equation}
such that as $r\to \infty$, $f(r) \to  r^2/L^2$ and $h(r)\to r/L$. The equations of motion then take the form:

\begin{subequations}\label{eq:eom}
\begin{align}
\label{eq:eom:1}\psi''(r) +\left(\frac{f'(r)}{f(r)}+2\frac{h'(r)}{h(r)}\right)\psi'(r)-\frac{V'(\psi)}{2f(r)}&=0\\
\label{eq:eom:2} h''(r) + \frac{\psi'(r)^2}{2}h(r)&=0\\
\label{eq:eom:3} f''(r) + 2\frac{f'(r)h'(r)}{h(r)}+V(\psi(r))&=0\\
\label{eq:eom:4} \frac{h'(r)^2}{h(r)}+\frac{f'(r)h'(r)}{f(r)}-\frac{h(r)(\psi'(r))^2}{2}+\frac{h(r)V(\psi)}{2f(r)}&=0
\end{align}
\end{subequations}
Since we want to consider the low temperature behavior of the conductivity, we are interested in solutions with a small black hole.  We will find such solutions numerically by imposing  boundary conditions of regularity on the horizon, 
the above asymptotic conditions on the metric, and a boundary condition for the scalar field which we discuss next.


\subsection{Double-trace boundary conditions}
\label{sec:multitrace}

We can deform our boundary CFT by adding the following double-trace operator to the boundary action
\begin{equation}
S\to S- \kappa \int d^3 x \ \mathcal{O}^2
\label{eq:deform}
\end{equation}
where  $\mathcal{O}$ is the operator dual to $\psi$.
This deformation is relevant if the dimension of  $\mathcal{O}$ is less than $3/2$. 
If $\kappa>0$, then this term increases the energy and makes it harder for $\mathcal{O}$ to condense.  However, if $\kappa<0$, we have the opposite behavior and there is some critical temperature $T_c$ below which  $\langle \mathcal{O}\rangle \neq 0$ \cite{Faulkner:2010gj}. One might have thought that taking $\kappa<0$ would destabilize the theory and cause it not to have a stable ground state. However this is not the case. It has been shown that the energy of the dual gravity solution is still bounded from below, provided $V$ can be derived from a suitable superpotential \cite{Faulkner:2010fh} (which is true for a large class of potentials including the one we have chosen).

Recall that the dimension of the operator $\mathcal{O}$ is related to the mass of the scalar field in the bulk:
\begin{equation}
\Delta_{\pm} = 3/2 \pm \sqrt{9/4 +m^2L^2}
\label{eq:scaling}
\end{equation}
and
\begin{equation}
\lim_{r\to \infty} \psi(r) = \frac{\alpha}{r^{\Delta_-}}+\frac{\beta}{r^{\Delta_+}}+...
\label{eq:psibdy}
\end{equation}
As long as 
\begin{equation}
-\frac{9}{4L^2} < m^2 < -\frac{5}{4L^2}
\end{equation}
both of these modes are normalizable. In order for the operator in \eqref{eq:deform} to be a relevant deformation, we must take $\mathcal{O}$ to have dimension $\Delta_-$.

Our double trace deformation induces the following boundary condition on $\psi$ \cite{Witten:2001ua,Berkooz:2002ug}:
\begin{equation}
\beta = \kappa\alpha
\end{equation}
Note that expanding our potential in \eqref{eq:gpsi} to second order in $\psi$ gives a mass $m^2 = -2/L^2$, within the range required by these boundary conditions. Note that this also tells us $\Delta_- = 1, \Delta_+ =2$, and 
\begin{equation}
\langle \mathcal{O} \rangle = \alpha
\end{equation}

To understand precisely how introducing a double trace deformation with $\kappa < 0$ can cause the Schwarzschild-AdS solution to become unstable at low temperature, we refer the reader to~\cite{Faulkner:2010gj}. 
We will just summarize an important point from that work motivating the existence of black hole solutions with nonzero scalar field below some critical temperature $T_c$.

 At finite temperature with no scalar field, the spacetime is described by planar AdS-Schwarzschild in $3+1$ dimensions,\footnote{From here on, we set $L=1$.}
\begin{equation}
d s^2 = -f(r)d t^2 + \frac{d r^2}{f(r)} + r^2 d \vec{x}^2 \quad\quad\text{where}\;\;f(r) = r^2\left (1- \frac{r_0^3}{r^3}\right )
\end{equation}
with a temperature, $T = 3r_0/4\pi$. We would like to find a condition for when the scalar field can be non-zero. At small values of $\psi$   our potential is approximately $V(\psi) \approx -6 - 2\psi^2+\mathcal{O}(\psi^4)$. Neglecting the higher order terms, we can exactly solve the scalar wave equation in the AdS-Schwarzschild background:  
\begin{equation}
\psi(r) = c_1 \left (\frac{r_0}{r}\right ) \;_2F_1\left [\frac{1}{3},\frac{1}{3},\frac{2}{3},\left (\frac{r_0}{r}\right )^3\right ] + c_2 \left (\frac{r_0}{r}\right )^{2}\; _2F_1\left [\frac{2}{3},\frac{2}{3},\frac{4}{3},\left (\frac{r_0}{r}\right )^3\right ]
\end{equation}
For this solution to be well behaved on the horizon, we need 
\begin{equation}\label{eq:reg}
\frac{c_2}{c_1} =  -\frac{\Gamma(2/3)^3}{\Gamma(4/3)\Gamma(1/3)^2}
\end{equation}
to cancel the diverging  logarithmic pieces from the hypergeometric functions. The large $r$ expansion of $\psi$ gives
\begin{equation}
\lim_{r\to\infty}\psi(r) = c_1 r_0/r + c_2(r_0/r)^{2}+...
\end{equation}
which, written in terms of the multitrace boundary condition gives,
\begin{equation}\label{eq:kappa}
\frac{\beta}{\alpha} = \kappa = \frac{c_2}{c_1}(4\pi T/3)
\end{equation}
Since $\kappa$ is negative and has the same dimensions as temperature, it is convenient to work with the (positive) dimensionless quantity $T/(-\kappa)$. 
Using (\ref{eq:reg}) and \eqn{eq:kappa}  one finds the critical value
 at which the static scalar field with double-trace boundary conditions is regular on the horizon:

\begin{equation}
\frac{T_c}{(-\kappa)} = \frac{3}{4\pi}\left(\ \frac{\Gamma(4/3)\Gamma(1/3)^2}{\Gamma(2/3)^3}\right)
\label{eq:kappa}
\end{equation}

\begin{figure}[tbp]
  \centering
    \includegraphics[width=0.6\textwidth]{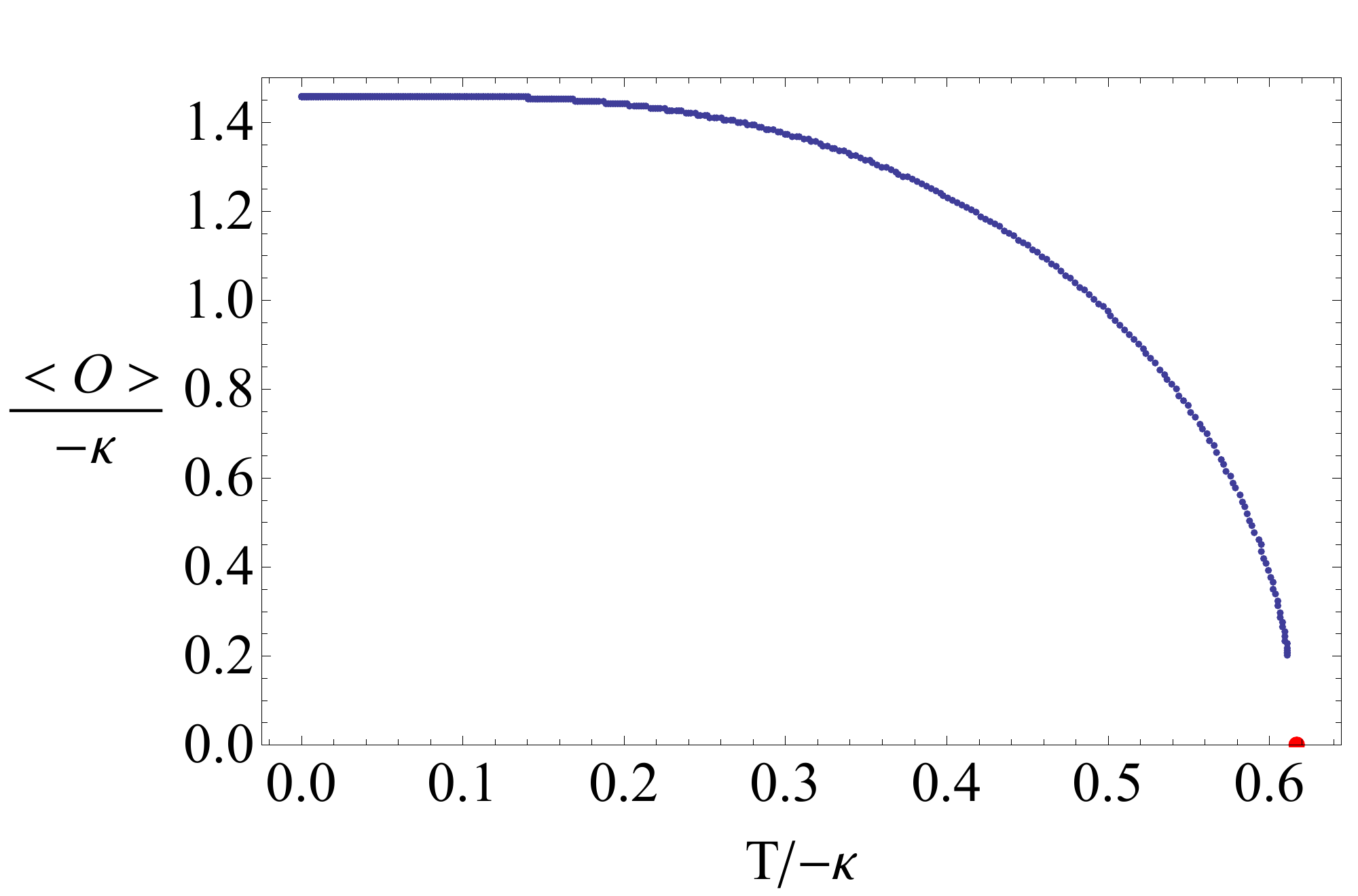}
  \caption{\label{fig:condensate} The value of $\langle\mathcal{O}\rangle$ vs. $T/(-\kappa)$ with critical value $T_c/(-\kappa) \approx .616$. }
  \end{figure}

This corresponds to a critical temperature of $T_c/(-\kappa) \approx .616$. 
So for any $\kappa < 0$, when $T = T_c$ there is a static linearized mode of the scalar field. This 
 signals the onset of an instability to forming scalar hair. At lower temperature, the scalar field is nonzero outside the black hole. From its asymptotic value, one finds that  $\langle \mathcal{O} \rangle$  increases as we lower $T$ and approaches a constant as $T\rightarrow 0$ (see Fig. \ref{fig:condensate}).
     
\subsection{Solutions}
\label{sec:domainwall}

Lowering the temperature (or equivalently, decreasing $\kappa$) below its critical value causes  the scalar field to roll down the potential $V(\psi)$. Since we have chosen $V(\psi)$ \eqref{eq:gpsi} to have a global minimum at $\psi_c$,  as $T \to 0 $ the value of the the scalar on the horizon approaches $\psi_c$. The zero temperature solution is  thus a renormalization group flow from an asymptotic $AdS_4$ as $r\to\infty$ to a new $AdS_4$ in the IR as $r\to0$ whose length scale is determined by the minimum value of the potential. Furthermore, the scalar field will have a new mass given by oscillations about this global minimum which governs its scaling dimension in the deep IR. Expanding about the global minimum of \eqref{eq:gpsi}, we see
\begin{equation}
V(\psi_c + \delta\psi) = -8 + 8\delta\psi^2+...
\end{equation}
Setting $V(\psi_c) = -6/L_{IR}^2$, so $L_{IR}$ is the AdS radius in the IR, we have   
$L_{IR}^2 = 3/4$. Using this, we find from \eqref{eq:scaling} that 
\begin{equation}
\Delta_{\pm}^{IR} = \frac{3}{2}\pm \sqrt{9/4+m^2_{IR}L^{2}_{IR}} = \frac{3\pm\sqrt{33}}{2}
\end{equation}
At zero temperature, the only normalizable solution in the IR scales like
\begin{equation}\label{eq:deltapsi}
\delta \psi(r) \equiv \psi_c  -\psi(r)  \sim r^{\Delta_\psi},  \ \ {\rm with}\ \  \Delta_\psi \equiv - \Delta^{IR}_- \approx 1.37228
\end{equation}

At very low temperature, the black hole horizon is at small $r_0$ where the scalar field is essentially constant  $\psi \approx \psi_c$. One thus expects that 
 the spacetime should look like planar AdS-Schwarzschild with the replacement $L^2 \to L_{IR}^2$. One also expects that the scalar field will not be modified much by the horizon, so that the value of the scalar field on the horizon will scale like 
 $\delta \psi(r) \sim r_0^{\Delta_\psi}$.

\begin{figure}[tbp]
  \centering
    \includegraphics[width=0.49\textwidth]{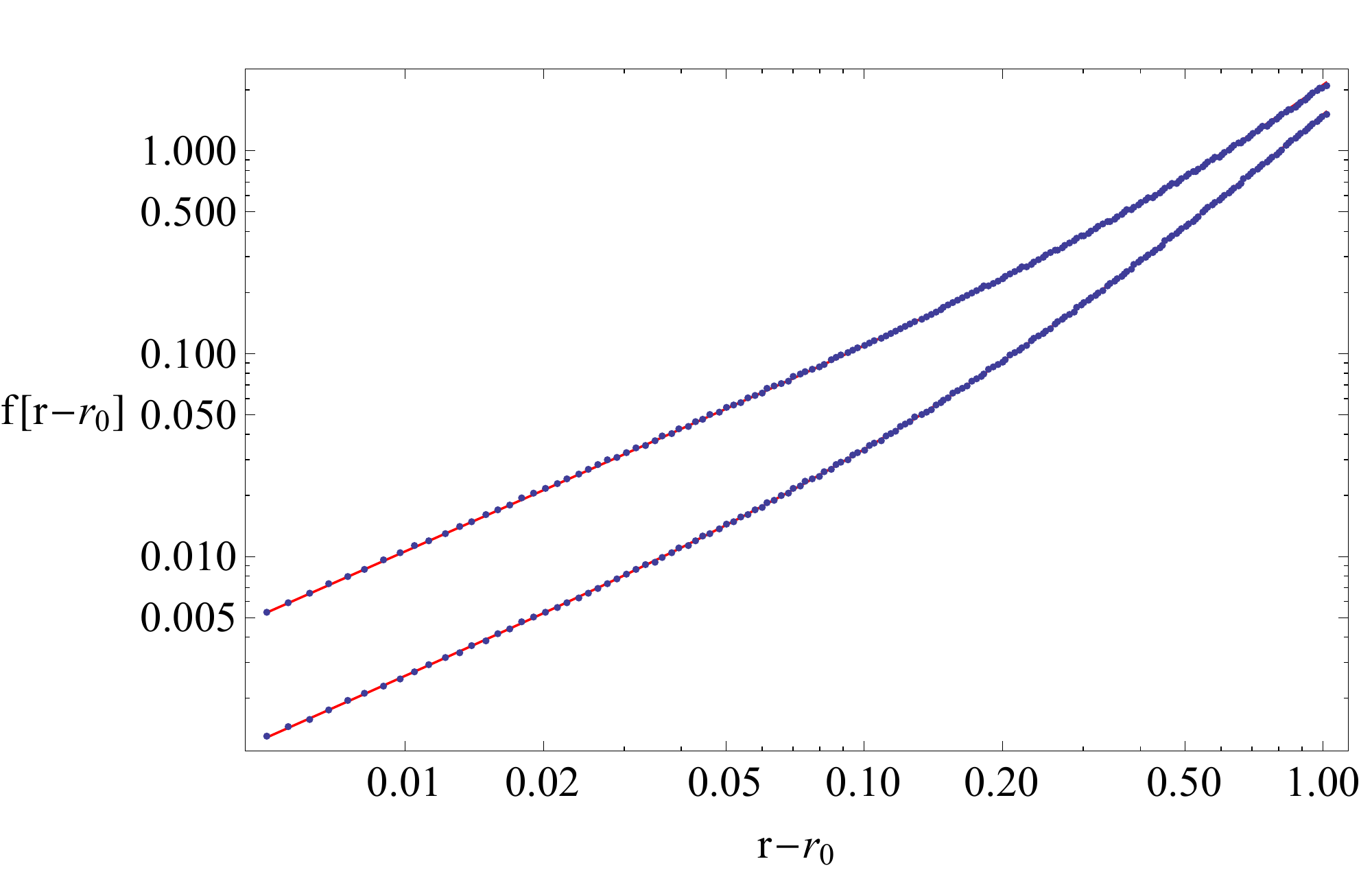}
\hfill
\includegraphics[width=0.49\textwidth]{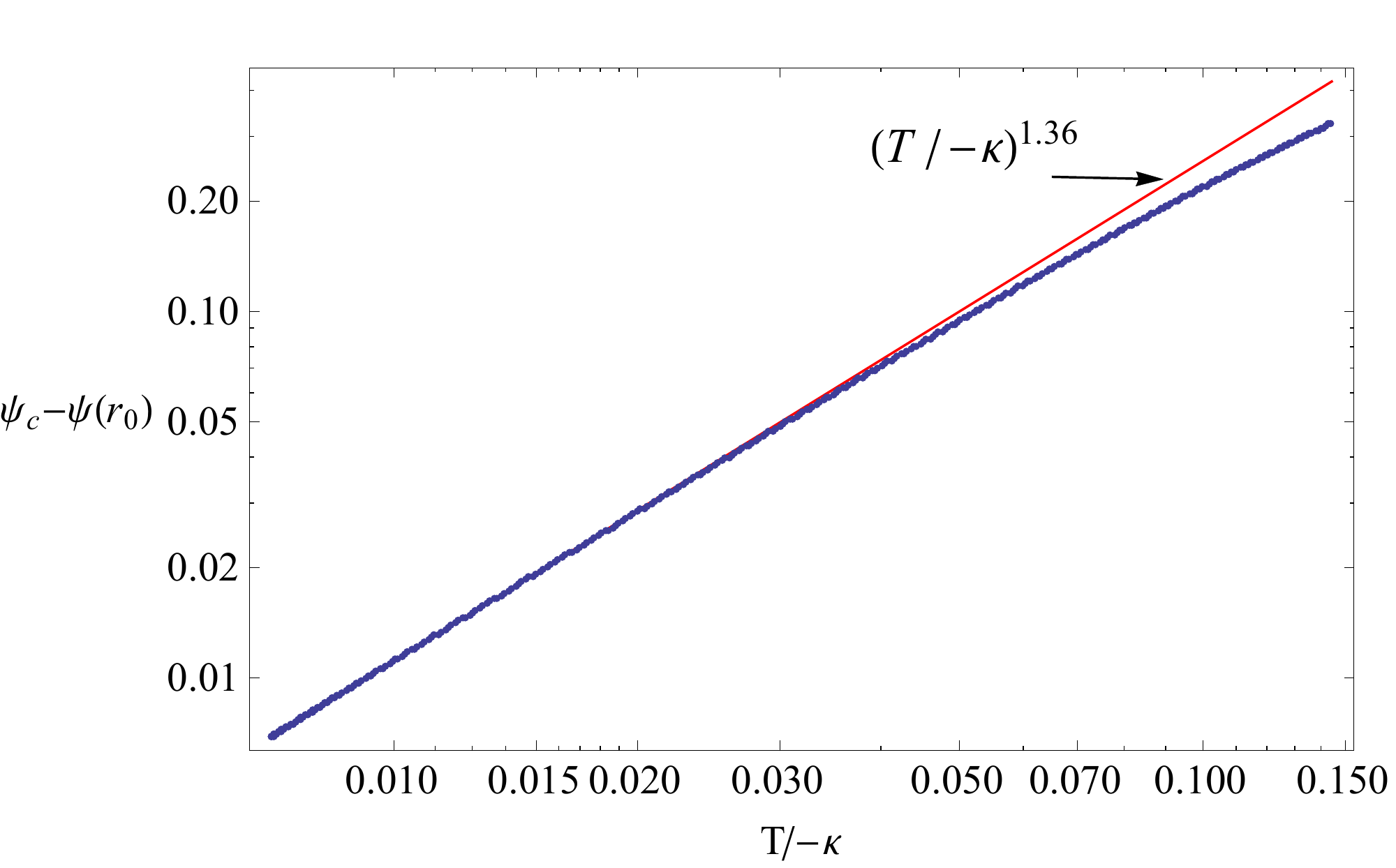}
  \caption{(Left) Log-log plot of our numerical solution for $f(r)$ in the IR (blue dots). The red line (which goes through all the points) is the analytic planar AdS-Schwarzschild solution. The two curves correspond to $T/(-\kappa) = .037$ (top) and  $T/(-\kappa) = 7.06 \times 10^{-3}$ (bottom).  One can see the transition from the linear Schwarzschild behavior to the quadratic  $AdS_4$ behavior. (Right) The scalar field on the horizon as a function of temperature. At low $T$, it scales like $T^{\Delta_\psi}$.}
 \label{fig:fapsi}
\end{figure}

 To check these expectations we solve the equations numerically. (See the end of the next section for a brief discussion of our numerical methods.) As shown in Fig.
 \ref{fig:fapsi} our results confirm these expections. On the left we show a plot of the metric function $f(r)$
and on the right is a plot of the scalar field evaluated on the horizon. The black hole  has a temperature $T = 3r_0/4\pi L_{IR}^2 = r_0/\pi$, and an entropy scaling like $S \sim T^2$.


\section{Conductivity}
\label{sec:conductivities}

Since our dual theory is conformally invariant in the IR, we would expect the conductivity to be characterized by power laws. We now demonstrate this is the case.  As usual, to calculate the conductivity, we perturb our spacetime with a harmonically time varying electric field. To do so, we introduce $\delta A_x = a_x(r) e^{-i\omega t}$. This perturbation back reacts to give at first order a metric perturbation $\delta g_{tx}$ with no other metric components being affected. Einstein's equation for this component of the metric and the equation of motion for $a_x$ give two coupled second order ODE's which can be combined to give the following equation for $a_x$,
\begin{equation}
a_x''(r) + \left(\frac{f'(r)}{f(r)} +\frac{ G'(\psi)\psi'(r)}{G(\psi)}\right)a_x'(r) + \frac{\omega^2}{f(r)^2}a_x(r) =0.
\label{eq:conductivity}
\end{equation}
We can solve this equation numerically using our background  solution subject to the boundary condition that the gauge field is ingoing at the horizon \cite{Son:2002sd}. The asymptotic behavior of $a_x$ is given by
\begin{equation}
\lim_{r\to\infty} a_x(r) = a_x^{(0)} + \frac{a_x^{(1)}}{r}+...
\end{equation}
When our perturbation corresponds to an applied electric field $E$ with harmonic time dependence, then $a^{(0)}_x = -iE/\omega$ and gauge/gravity duality implies~\cite{Hartnoll:2009sz}  $a^{(1)}_x = \langle J \rangle$ so that our conductivity is given by
\begin{equation}
\sigma(\omega) = \frac{a^{(1)}_x}{i\omega a^{(0)}_x}.
\end{equation}

\subsection{DC conductivity}
\label{sec:dcconductivity}

As first realized by Iqbal and Liu~\cite{Iqbal:2008by}, low frequency limits of transport coefficients in the dual field theory are determined by the horizon geometry of the gravity dual. This is a holographic application of the ``membrane paradigm" of classical black holes. Applied to a $U(1)$ gauge field, this implies that the DC conductivity is given by the coefficient of the gauge field kinetic term evaluated on the horizon. To see this, assume $T>0$ and  consider the $\omega\ll T$ limit of eq. \eqref{eq:conductivity}. In this limit, the last term can be neglected\footnote{Even though this term diverges at the horizon, at nonzero temperature $f(r)$ vanishes linearly, so the horizon is a regular singular point of \eqref{eq:conductivity}. This is no longer true when $T=0$.} so that the equation can be rewritten
\begin{equation}\label{eq:dc}
\frac{1}{f(r)G(\psi)}[f(r)G(\psi)a_x'(r)]' = 0
\end{equation}
Now, on the  boundary, the conserved quantity in this equation becomes
\begin{equation}
\lim_{r\to \infty} f(r)G(\psi)a_x'(r) = r^2\left(\frac{-\langle J \rangle}{r^2}\right) = -\langle J \rangle
\end{equation}
Normally, we would have to solve eq. \eqref{eq:conductivity} numerically to find $\langle J \rangle$. However, because this is conserved in the DC limit, we can evaluate it on the horizon.
\begin{equation}
\lim_{r\to r_0} f(r)G(\psi)a_x'(r) = (1+g\psi^2)^2 f(r)a_x'(r)|_{r=r_0}
\label{eq:conserved}
\end{equation}
Now, our ingoing boundary conditions  tell us that on the horizon, $a_x$ must be a function of the tortoise coordinate, $dr_* = dr/f(r)$ in the combination $v = t+r_*$\footnote{This is of course only valid at nonzero $\omega$ where we have harmonic time dependence. We compute the low frequency conductivity and then take $\omega\to 0$.}. This allows us to relate time derivatives to radial derivatives,
\begin{equation}
\partial_t a_x(u) |_{r=r_0} = \partial_v a_x |_{r=r_0} = f(r)\partial_r a_x|_{r=r_0}
\end{equation}
so that eq. \eqref{eq:conserved} becomes
\begin{equation}
\langle J \rangle = -(1+g\psi(r_0)^2)^2 \partial_t a_x(r)|_{r=r_0}
\end{equation}
In the low frequency limit, $dF = 0$ implies that the electric field is essentially constant. We can thus evaluate it near  the horizon and see that
\begin{equation}
\sigma_{DC} =\lim_{\omega\to 0} \sigma(\omega) = \frac{i\omega a_x(r_0) (1+g\psi(r_0))^2}{i\omega a_x(r_0)} = [1+g\psi(r_0)^2]^2
\end{equation}
If we choose $g=-1/\psi_c^2$, then at very low temperatures, with our AdS-Schwarzschild domain wall solution \eqn{eq:deltapsi}, we have
\begin{equation}
\sigma_{DC} \sim (\delta\psi(r_0)/\psi_c)^2 \sim r_0^{2\Delta_{\psi}} \sim T^{2\Delta_{\psi}}
\label{eq:DCcond}
\end{equation}


\subsection{Optical conductivity}
\label{sec:optical}

We now investigate how the choice of $g = -1/\psi_c^2$ affects the zero temperature optical conductivity. At zero temperature, we have purely $AdS_4$ in the IR part of the domain wall. In this background, equation \eqref{eq:conductivity} can be solved exactly to give,
\begin{equation}
a_x(r) \sim  i(\omega L_{IR}^2 r)^{-(1+2\Delta_\psi)/2} H^{(1)}_{\frac{1+2\Delta_\psi}{2}}\left (\frac{\omega L_{IR}^2}{r}\right )
\label{eq:asolution}
\end{equation}
where we have written the solution in terms of a Hankel function, such that as we approach the Poincar\'e horizon in the IR, 
\begin{equation}
H^{(1)}_{\frac{1+2\Delta_\psi}{2}}\left (\frac{\omega L_{IR}^2}{r} \right ) \sim \sqrt{\frac{\pi r}{\omega L_{IR}^2}}e^{i\omega L_{IR}^2/r } =  i\sqrt{\frac{\pi}{r_*}}e^{-i\omega r_*}
\end{equation}
as required. Now, it is worth pointing out two important features of this solution. The first is that, because we have a domain wall,  this solution only holds for $r<r_D$, where $r_D$ is the location of the domain wall, which must be found numerically. The second point is that the solution above is the zero temperature solution. But we have seen that at low temperature, in the region $r_0 \ll  r  \ll r_D$ the spacetime is essentially $AdS_4$, and we have checked numerically that the above solution is still  valid.

\begin{figure}[tbp]
  \centering
\includegraphics[width=0.7\textwidth]{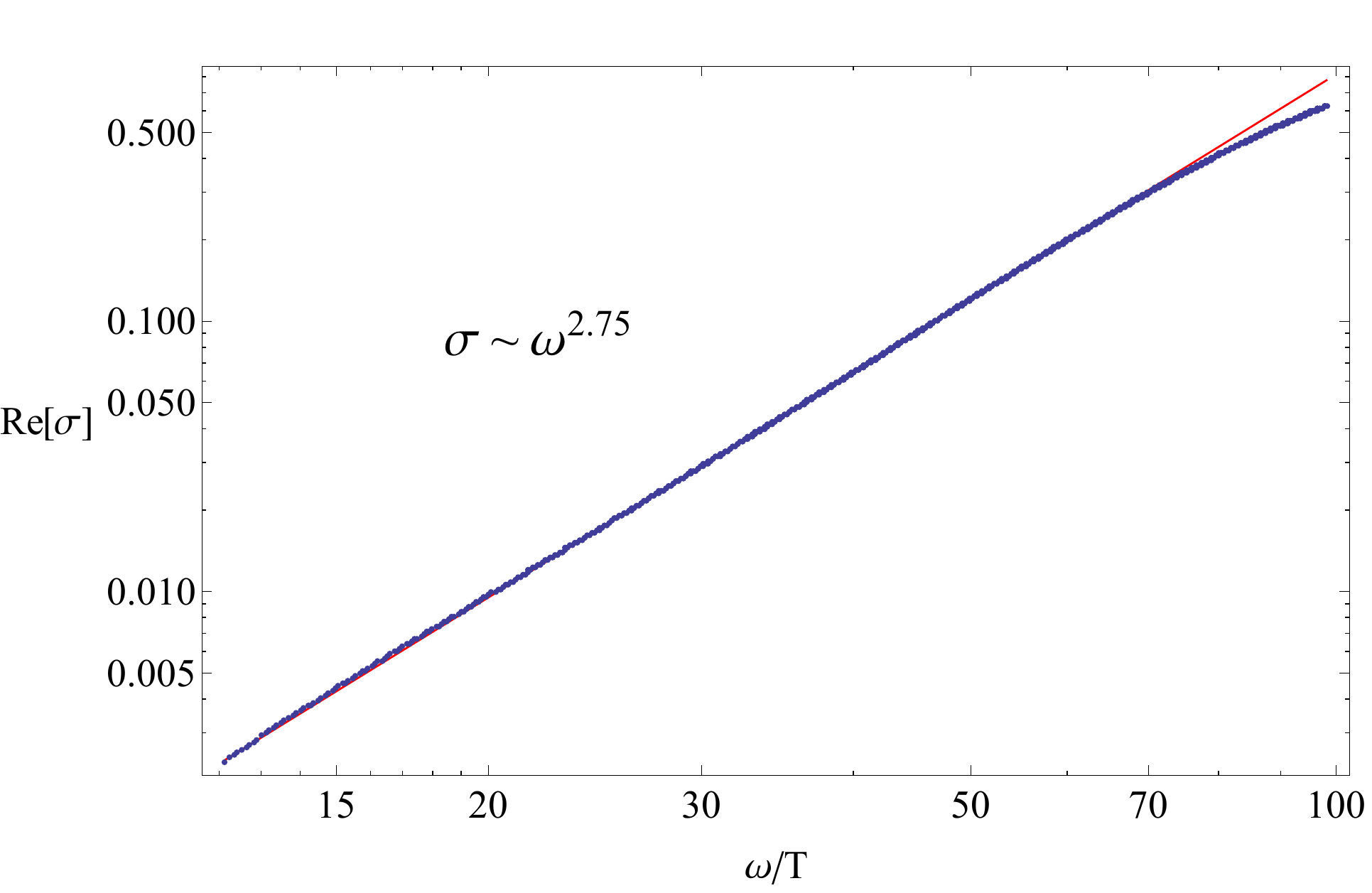}
  \caption{ Log-log plot of the optical conductivity vs. frequency at $T/(-\kappa) = 7.06\times 10^{-3}$. The line of best fit gives $\sigma \sim \omega^{2.75}$. Our analytic solution says that it should be a power law with an exponent $2\Delta_\psi \approx 2.744$.}
 \label{fig:fofr}
\end{figure}

To calculate the optical conductivity, we will use the matched asymptotic expansion of Gubser and Rocha~\cite{Gubser:2008wz}. The basis of this analysis rests on the presence of a conserved flux,
\begin{equation}
\mathcal{F} = -f G(\psi)a_x^*\overleftrightarrow{\partial_r}a_x.
\end{equation}
One can check that $\partial_r \mathcal{F}$ does indeed vanish by using the equation of motion \eqref{eq:conductivity}. In the UV, this conserved flux gives
\begin{equation}
\lim_{r\to\infty} \mathcal{F} = -a_x^{(0)}a_x^{(1)*} + a_x^{(0)*}a_x^{(1)}
\end{equation}
From this we see that we can calculate the real part of the conductivity
\begin{equation}
Re[\sigma(\omega)] = \lim_{r\to\infty} \frac{\mathcal{F}}{2i \omega |a_x^{(0)}|^2}.
\end{equation}
To determine this analytically, we need to find $a_x^{(0)}$. This is possible because we note that  in the DC limit (at low temperature), \eqref{eq:dc} allows for $a_x(r)$ to have a constant piece which is undetermined by the equation of motion. However, the low frequency limit of \eqref{eq:asolution} should smoothly match onto the DC solution and horizon boundary conditions allow us to fix this constant. A general solution of the DC equation \eqref{eq:dc} in the region $r_0 \ll  r  \ll r_D$ has the form
\begin{equation}
a_x(r) = Cr^{-(1+2\Delta_\psi)} + D
\end{equation}
  Expanding \eqref{eq:asolution} for small $\omega L_{IR}^2/r$, we see 
\begin{equation} 
a_x(r) \sim (\omega L_{IR}^2)^{-(1+2\Delta_\psi)}\left[ \frac{(i+\tan(\Delta_\psi \pi))}{\Gamma(\frac{3+2\Delta_\psi}{2})2^{(1+2\Delta_\psi)/2}}(\frac{\omega L_{IR}^2}{r})^{1+2\Delta_\psi} + \frac{2^{(1+2\Delta_\psi)/2}}{\pi}\Gamma(\frac{1+2\Delta_\psi}{2}) \right]
\label{eq:expansion}
\end{equation}
where we have pulled out an $\omega$ dependence as an overall normalization.  The expression inside the brackets is matched to the DC solution at low temperature.  The second piece corresponds to the D term. Because it has no $r$ dependence, its value in the IR part of the domain wall must match the value in the UV which is $a_x^{(0)}$. We can then use \eqref{eq:expansion} to evaluate the conserved flux in a region $r_0 \ll  r  \ll r_D$. Finally, because the real part of the conductivity is the ratio of two conserved quantities, evaluating these quantites in this region is equivalent to computing the conductivity on the boundary. Doing so gives a power law in the low temperature optical conductivity, 
\begin{equation}
\lim_{T\to 0} Re[\sigma]\sim \frac{i\omega^{-(1+2\Delta_\psi)}}{i\omega(\omega^{-2(1+2\Delta_\psi)})} =  \omega^{2\Delta_\psi}.
\end{equation}
This behavior is confirmed by our numerical solutions as shown in Fig. \ref{fig:fofr}.


\subsection{Comment on numerical methods}

The equations of motion \eqref{eq:eom} and \eqref{eq:conductivity} are second order differential equations. At non-zero temperatures, these lend themselves well to pseudospectral methods \cite{Headrick:2009pv, Figueras:2011va}. It is well-known that low temperature numerics are difficult to study numerically because as $T\to 0$, the metric function $f(r)$ vanishes quadratically. For this reason, we found that for low temperatures, we needed a 400 point Chebyshev grid to minimize numerical noise and optimize precision in computing the conductivity. For our pseudospectral methods to cover the full spacetime, we used a variable $z\equiv 1/r$ and rescaled the horizon to $r_0 = 1$. After solving the equations of motion, we rescaled the horizon back to the proper $r_0 = \pi T$. Furthermore, because we fixed $r_0$, our temperature was varied by adjusting the parameter $\kappa$ in the boundary conditions for our scalar field. All data showing the temperature dependence is plotted in terms of the dimensionless quantity $T/(-\kappa)$. Finally, we rescaled our functions $f(z) \to F(z)/z^2$ and $h(z) \to H(z)/z$ to be well behaved at the conformal boundary $z\to0$. We also rescaled the gauge field $a_x(z) \to e^{-i\omega z_*}A_x(z)$ to be better behaved on the horizon. The appropriate boundary condition for the redefined functions are $F'(0) = H'(0) = 0$ on the boundary and $F(1)A_x'(1) = 0$ corresponding to ingoing boundary conditions at the horizon. 


\section{Discussion }
\label{sec:discussion}

We have presented a nonsingular holographic model of an insulator. A key parameter in the model, $g$,  controls the 
 coupling between the kinetic term for the gauge field and a neutral scalar field. A scalar potential with a global minimum at $\psi =\psi_c$ allows us to define a critical $g_c = -1/\psi_c^2$ such that the dual theory has a DC conductivity that goes to zero as a power of the temperature $T$. This same critical $g_c$ produces a zero temperature optical conductivity that also vanishes as a power of $\omega$. Both exponents agree and are given by the scaling dimension of the scalar field in the IR.
 This behavior  has also been seen in models with nonzero charge density and broken translation invariance ~\cite{Donos:2012ra,Hartnoll:2012rj,Donos:2013eha,Donos:2014uba}.

\begin{figure}[tbp]
  \centering
    \includegraphics[width=0.7\textwidth]{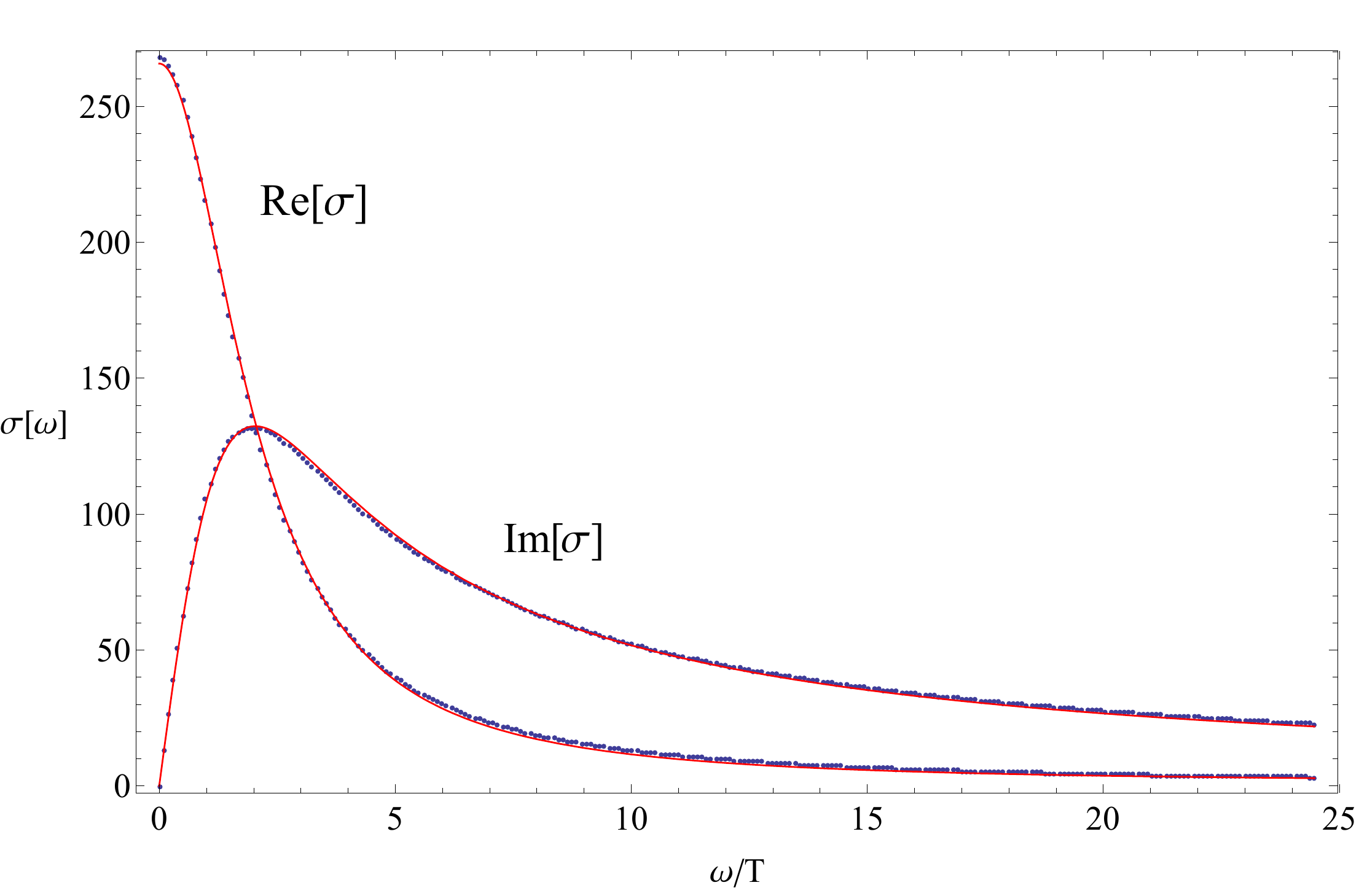}
  \caption{\label{fig:drude}Fit of optical conductivity to a Drude type curve, $\sigma(\omega) = \frac{K\tau}{1-i\omega\tau}$. For this plot, we chose $g=10$ at a temperature $T/(-\kappa) = .037$, and found $K = 540$, $\tau = .485$.}
\end{figure}

We now ask what happens for other values of $g$.
 We know that $g \to g_c$ effectively increases the interactions between the charge carriers causing  $\sigma_{DC}\to 0$. As we increase $g$, $ \sigma_{DC}$ also increases. It reaches one when $g=0$, which is expected since this is the standard value for the conductivity in AdS-Schwarzschild, and $g=0$ turns off the coupling between the scalar and gauge field. For $g >0$, $ \sigma_{DC} > 1$. For large $g$ there is a pronounced Drude peak showing that we have a standard metal.  This is illustrated in Fig. \ref{fig:drude}.

\begin{figure}[tbp]
  \centering
    \includegraphics[width=0.4\textwidth]{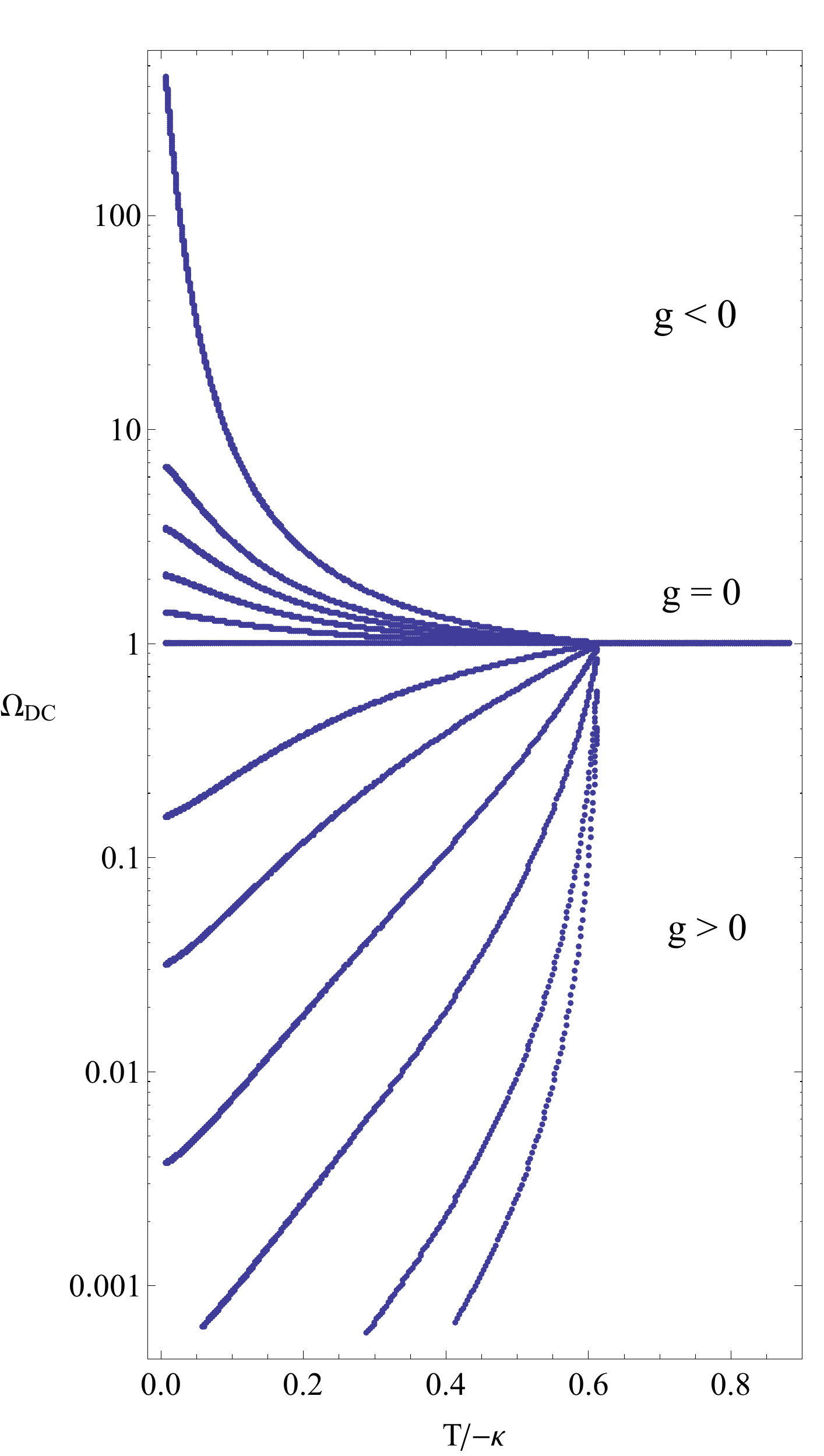}
\hfill
\includegraphics[width=0.53\textwidth]{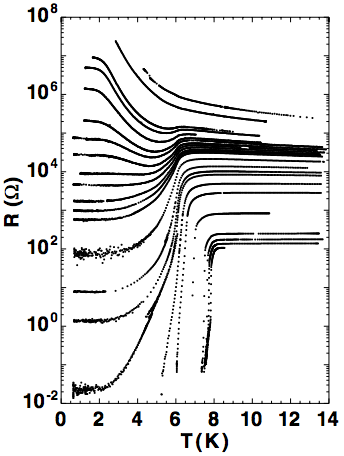}
  \caption{\label{fig:bosemetal}(Left) Plot of  DC resistivity vs. temperature for different values of our tuning parameter $-1/\psi_c^2 < g < 200$.  (Right) Numerical data from ~\cite{y} showing the DC resistivity of a Ga thin film. The resistivity increases as the thickness of the thin film is decreased (bottom to top).}
\end{figure}

In Fig. \ref{fig:bosemetal} on the left, we have plotted our numerical results for the DC resistivity, $\Omega \equiv 1/\sigma_{DC}$, as a function of temperature for different values of $g\geq g_c$. 
Just for fun, on the right is experimental data from a Bose metal~\cite{u}. Bose metallicity is a unique phase exhibited by certain thin film materials which also exhibit high temperature superconductivity. These materials are characterized by strong interactions among their charge carrying quasiparticles and conductivity along two-dimensional planes. By applying a magnetic field transverse to these planes or by adjusting the thickness of the thin films, one can create a phase where Cooper pairs (bosons) have condensed but the global $U(1)$ symmetry has not been broken.\footnote{This effect is unique to two (spatial) dimensions where phase coherence fall-offs are algebraic, $G(r)\sim r^{-\eta}$ with $0<\eta<1$~\cite{u}} The lack of phase coherence gives a finite DC conductivity, hence the name Bose metal. 

The two sets of curves in Fig. \ref{fig:bosemetal} show an interesting similarity. However, the experimental curves on the right are obtained by increasing the thickness of a thin film while on the left we are changing a parameter in the bulk Lagrangian and therefore modifying the $2+1$ boundary theory. To better describe a Bose metal, we would need to tune a parameter in the boundary theory instead of the bulk. This could be done by introducing a new bulk field which couples to $\psi$ and effectively modifies its potential to have a new minimum at  $\psi_c' < \psi_c$. Certain lattice models \cite{Donos:2013eha,Donos:2014uba} seem capable of such a deformation.


\acknowledgments
We would like to thank G. Hartnett and B. Way  for help with numerics. This work was supported in part by NSF grant PHY12-05500.


\end{document}